\documentclass[journal]{IEEEtran}
\IEEEoverridecommandlockouts
\usepackage{cite}
\usepackage{amsmath,amssymb,amsfonts}
\usepackage{algorithmic}
\usepackage{graphicx}
\usepackage{textcomp}
\usepackage{xcolor}
\usepackage{soul}
\def\BibTeX{{\rm B\kern-.05em{\sc i\kern-.025em b}\kern-.08em
    T\kern-.1667em\lower.7ex\hbox{E}\kern-.125emX}}

\begin{document}

%\title{Advancing Node-Specific Nerve Regeneration: A New Wearable-to-Implant Wireless Power Transfer System Through Galvanic Body-Coupled Powering}
\title{Enhancing Wireless Power Transfer in Neural Microdevices: The Role of Galvanic Body-Coupled Powering with Circular Ring Transmitters}
\title{Efficient Galvanic Body-Coupled Powering for Wireless Neural Microdevices}
\title{Efficient Galvanic Body-Coupled Powering for Wireless Implanted Neurostimulators}
\title{Galvanic Body-Coupled Powering\\for Wireless Implanted Neurostimulators}

% {\footnotesize \textsuperscript{*}Note: Sub-titles are not captured in Xplore and
% should not be used}
% \thanks{Identify applicable funding agency here. If none, delete this.}
% }
\author{\IEEEauthorblockN{Asif Iftekhar Omi\textsuperscript{1}, Emma Farina\textsuperscript{2}, Anyu Jiang\textsuperscript{1}, Adam Khalifa\textsuperscript{1}, Shriya Srinivasan\textsuperscript{2}, and Baibhab Chatterjee\textsuperscript{1}}
\IEEEauthorblockA{\textit{\textsuperscript{1}Department of Electrical and Computer Engineering, University of Florida,}
Gainesville, USA.\\
\textit{\textsuperscript{2}Department of Bioengineering, Harvard University,}
Boston, USA.\\
email: \{as.omi, chatterjee.b\}@ufl.edu}
}
\maketitle
\vspace{-3mm}
\begin{abstract}
Body-coupled powering (BCP) is an innovative wireless power transfer (WPT) technique, recently explored for its potential to deliver power to cutting-edge biomedical implants such as nerve/muscle stimulators. This paper demonstrates the efficient technique of designing WPT systems embedding BCP via galvanic coupling (G-BCP). The G-BCP configuration utilizes two metal circular rings surrounding the body area of interest as the transmitter (TX) electrodes required for galvanic (differential) excitation and a wireless implant as the receiver (RX) equipped with two electrodes for differential power reception accordingly. By focusing on the unique advantages of this approach—such as enhanced targeting accuracy, improved power transfer efficiency (PTE), and favorable tissue penetration characteristics, G-BCP emerges as a superior alternative to traditional WPT methods. A comprehensive analysis is conducted to obtain the optimized device parameters while simultaneously allowing flexible placement of implants at different depths and alignments. To substantiate the proposed design concept, a prototype was simulated in Ansys HFSS, employing a multi-layered tissue medium of 10mm radius and targeting the sciatic nerve of a rat. Impressively, this prototype achieves $>$ 20\% PTE at 1.25 GHz, with the implant (radius of RX electrodes = 1 mm) located 2 mm deep inside the tissue model having complex load impedance of $R_{load} = 1000\Omega$ and $C_{load} = 5pF$. Therefore, the G-BCP-based wirelessly powered microdevices are envisaged to be a key enabler in neural recording and stimulation, specifically for the peripheral nervous system, enhancing therapeutic outcomes and patient experiences.
\end{abstract}
\vspace{-5mm}
\section{Introduction}
Implantable bioelectronics is a marvel in the contemporary healthcare sector as it continues to offer groundbreaking solutions that markedly improve the management of chronic diseases and refine surgical interventions. Devices such as pacemakers, retinal implants, and neural stimulators have become integral, providing targeted treatments \cite{chatterjee2023bioelectronic,agarwal2017wireless,barbruni2020miniaturised}. Particularly, wireless neural microdevices (WNMD) focus on delivering electrical impulses to nerve tissues to treat different neurological disorders or restore pertinent functions without the constraints of wired connections, frequent battery replacement, or cumbersome implants \cite{lee2021neural,chatterjee2023biphasic, khalifa2019microbead,Chatterjee_VLSI}. 

\begin{figure}[t]
\centerline{\includegraphics[width=0.91\columnwidth]{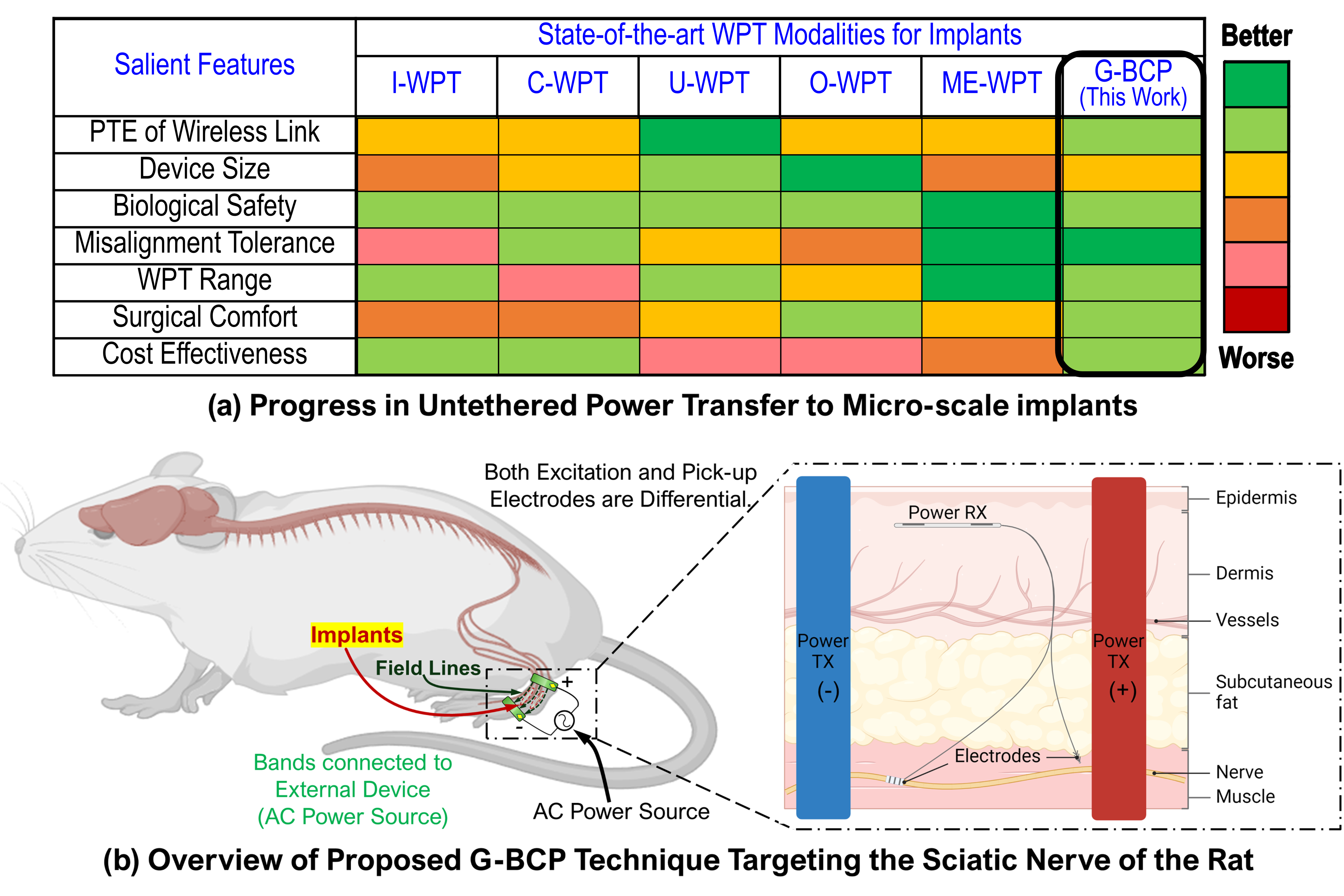}}
\vspace{-2mm}
\caption{Synopsis of different WPT techniques for bio-implants: (a) comparison of the salient features among state-of-the-art methods versus the proposed approach (b) conceptualization of the proposed G-BCP technique with the corresponding tissue layers. Here, the envisioned device comprises (1) a TX component with two circular \textcolor{black}{rings} for power transfer attached to the skin over the area of interest, (2) an RX component with two electrodes implanted subcutaneously, and (3) electrodes implanted within multi-layered tissue near or onto the nerve of interest, connected to the RX via leads.}
\label{intro1}
\vspace{-4mm}
\end{figure}

Despite these pivotal advances, effectively powering these miniaturized devices has remained one of the critical challenges within the biomedical circuits and systems community since the last decade \cite{monti2021wireless}. Traditional WPT methods incorporate near-field resonant inductive (I-WPT), capacitive (C-WPT), ultrasonic (U-WPT), optical (O-WPT) and magneto-electric (ME-WPT) techniques \cite{omi2024new, omi2024embc, koruprolu2018capacitive, ghanbari2019sub, moon2021bridging, alrashdan2021wearable, hosur2023comparative}. Overall, I-WPT is considered the paragon among all these state-of-the-art approaches highlighted in Fig.~\ref{intro1}(a), though it still fails to accurately target specific implants and maintain consistent power delivery through the complex medium of biological tissues \cite{monti2021wireless,roy2022powering,7015638,hosur2023comparative}. Possible misalignments further hinder power transfer, making these systems vulnerable to body movement, implant migration, and other external factors.

Recognizing these limitations, Body-Coupled Powering (BCP) is emerging as a promising area of research \cite{zhu2023biomedical, Modak_JSSC, chen2023intra}. BCP leverages the conductive properties of body tissues as channels for power transfer, enhancing efficiency and reducing tissue absorption compared to conventional wireless transmission systems. Although there is not a plethora of studies in the current literature on BCP, most of these works only demonstrate wearable-to-wearable scenarios, whereas the use of BCP for wearable-to-implants is still in the early stages of exploration\cite{han2023enhanced, han2021body, lee2022miniaturized,li2020body,li202034,li2021body,dong2021body,cho2022intra,tudela2021volume,minguillon2022powering,garcia2022wireless}. This paper delves into the innovative utilization of Galvanic BCP (G-BCP), as demonstrated in Fig.~\ref{intro1}(b). Unlike conventional methods that suffer from issues of power leakage and inefficiency, G-BCP ensures that power is precisely delivered to the intended node, thereby maximizing the functional efficacy of the implant and minimizing potential tissue damage. We show that a G-BCP power receiver implanted at a small depth is much less prone to misalignment as it can achieve $\approx$6$\times$ higher PTE than inductive coils due to \textcolor{black}{electric field-based coupling}. The power, once received, can be used for stimulating the nerve with a conventional tethered electrode (Fig.~\ref{intro1}(b)).  The key contributions of this proposed work are as follows:
\begin{itemize}
\item
Devising a new G-BCP system transferring power from skin-attached circular ring electrodes to the WNMD
% \vspace{-0.5mm}
\end{itemize}
% \vspace{-2mm}
\begin{itemize}
\item
Analyzing the trade-off among WNMD size, implant depth, resonant frequency and optimal power transfer
% \vspace{-1mm}
\end{itemize}
% \vspace{-2mm}
\begin{itemize}
\item
Maximizing PTE for any generic G-BCP model
% \vspace{-1mm}
\end{itemize}

Although the emphasis in this paper is on efficiently powering WNMD, which stimulates the sciatic nerve of a rat, the directions of the proposed methodology are broadly adaptable to other medical contexts, such as stimulating the human vagus nerve, spinal cord, or peripheral nerves.

\section{G-BCP Design Process}
\begin{figure}[t]
\centerline{\includegraphics[width=0.85\columnwidth]{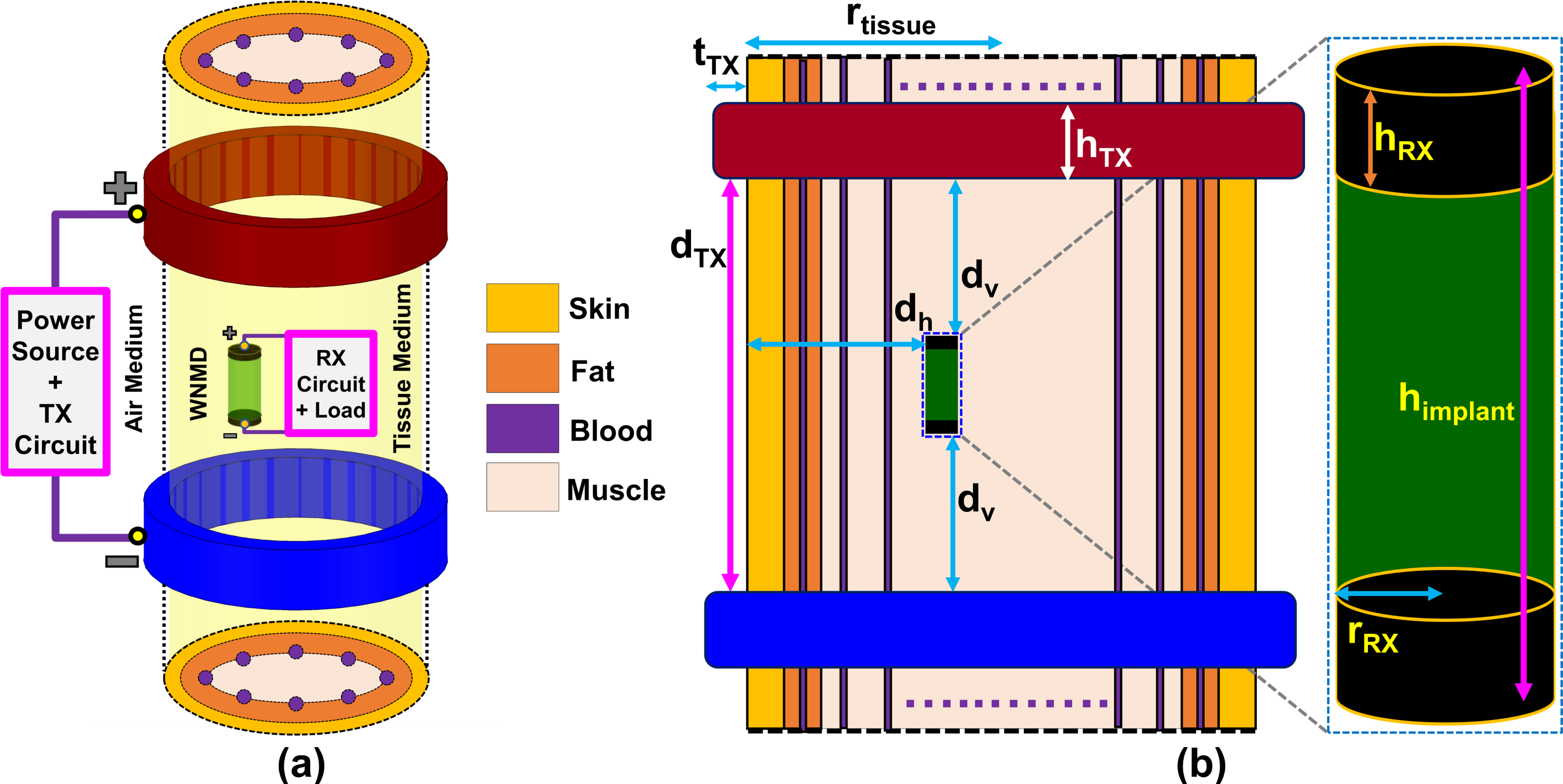}}
\vspace{-3mm}
\caption{Simplified G-BCP structure: (a) An overall system highlighting the TX ring electrodes and implanted WNMD. (b) Denoting the physical parameters. The TX and RX circuitry ensures power transfer from the source in the air to tissue load, whereas the G-BCP link maximizes PTE using galvanic excitation through the biological layers such as skin, fat, blood, and muscle.  Overall, analyzing the design parameters can give an idea about the available PTE, which is a key factor in designing the TRX circuits.}
\label{model}
\vspace{-5mm}
\end{figure}

With the G-BCP system depicted in Fig.~\ref{model}, we investigate the impact of various design parameters upon applying differential excitation to the TX electrodes and then provide a concise analysis aimed at optimizing PTE in this wireless link under arbitrary alignment of WNMD where the power is differentially coupled to RX electrodes.

\subsection{Device Modeling and Analysis}
Traditionally, galvanic electrodes are positioned laterally, aligning the implant depth with the distance between TX and RX electrodes \cite{pola2023galvanic}. In contrast, our design incorporates circular rings as TX electrodes encircling the targeted tissue area, affording greater flexibility in electrode placement and proximity to the implants.
\begin{figure}[t!]
\centerline{\includegraphics[width=1\columnwidth]{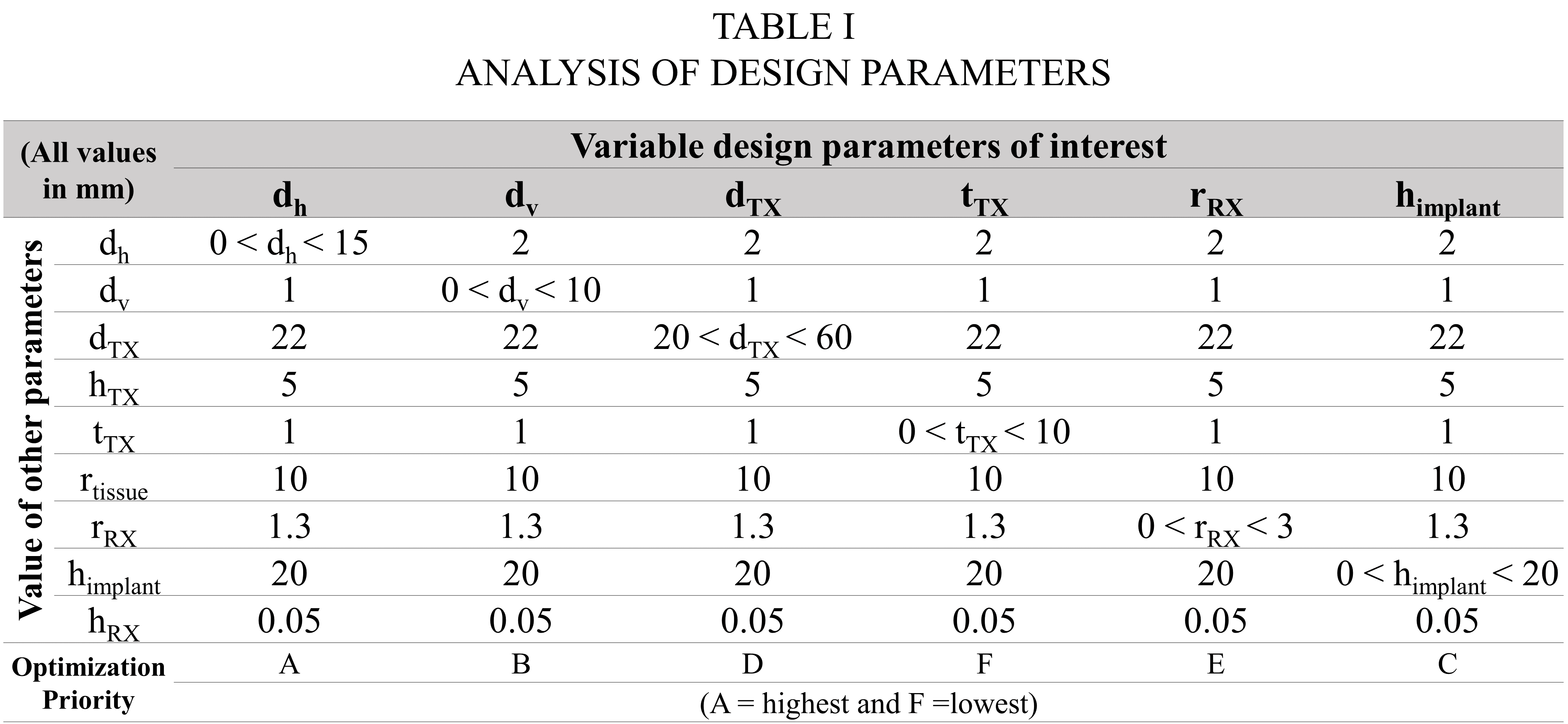}}
% \caption{G-BCP Model}
\label{u5}
\vspace{-5mm}
\end{figure}
\begin{figure}[htbp]
\centerline{\includegraphics[width=1\columnwidth]{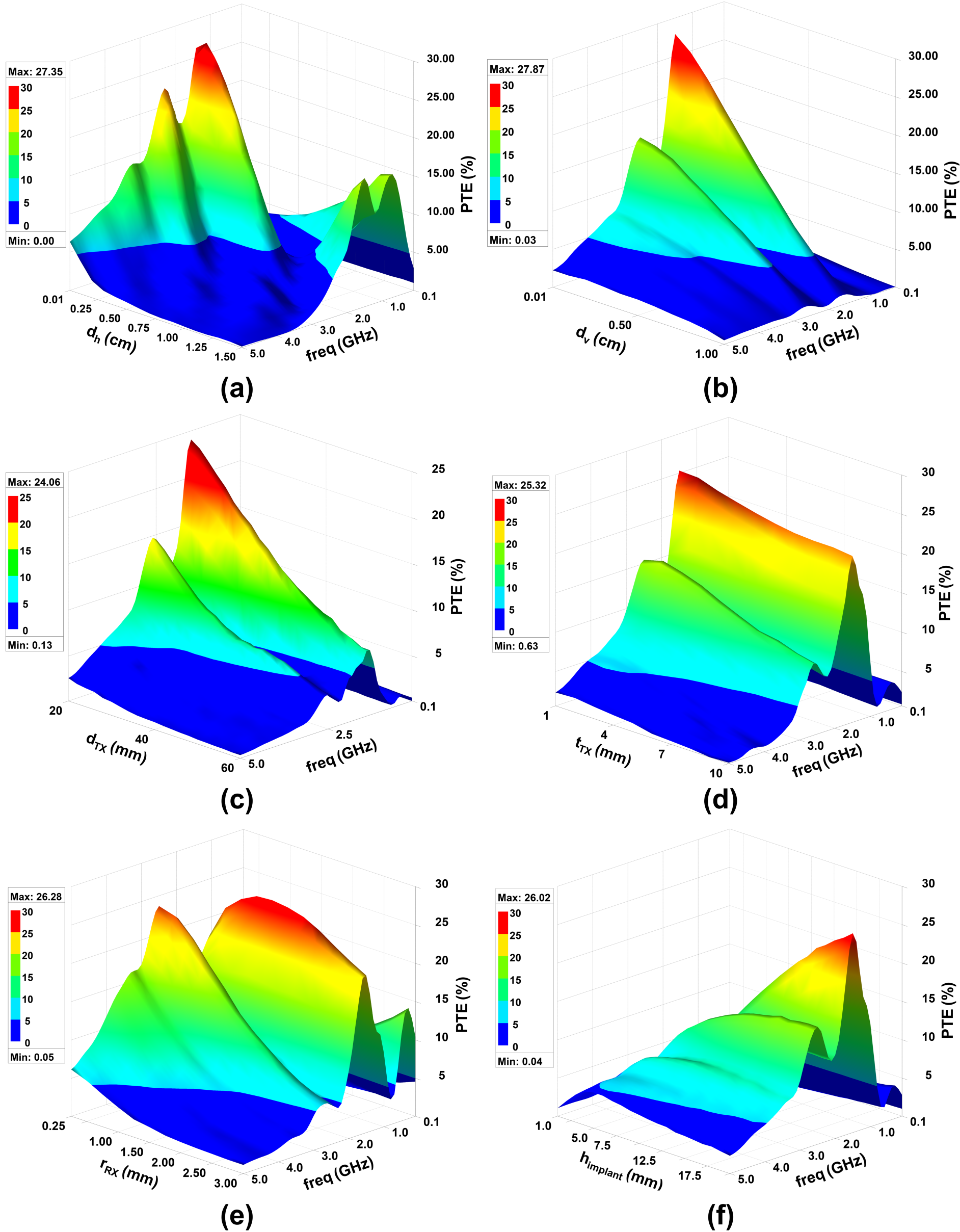}}
\vspace{-3mm}
\caption{Design space exploration: (a) impact of $d_{h}$ (b) impact of $d_{v}$ (c) impact of $t_{TX}$ (d) impact of $d_{TX}$ (e) impact of $r_{RX}$ (f) impact of $h_{implant}$. EM simulations are conducted using Ansys HFSS, an industry-standard FEM-based solver, with muscle tissue serving as the WPT medium, which is characterized by the Gabriel-Gabriel model for tissue dielectric properties \cite{gabriel1996dielectric}. The corresponding PTEs are obtained according to PTE ($\%$) = $100\times$ $\left | S_{21} \right |^{2}$ in matched conditions. For optimal performance, it is recommended to position TX rings closer to the implant and the RX implant nearer to the skin surface, particularly with an increased implant length.}
\label{variable_sweep}
\vspace{-5mm}
\end{figure}
Design parameters specified for clarity include implant depth ($d_{h}$), the vertical distance between TX and RX ($d_{v}$), the distance between circular TX electrodes ($d_{TX}$), the thickness of each TX electrode ($h_{TX}$), the distance between the outer and inner radius of the circular rings ($t_{TX}$), radius of the tissue medium ($r_{tissue}$), the radius of the RX electrodes ($r_{RX}$), the thickness of each RX electrode ($h_{RX}$), and overall length of the implant ($h_{implant}$), similar to \cite{Chatterjee_Arxiv_2022}. Table I depicts the range of these parameters, indicating the tuning priority of the variables in the design process, which is accelerated by adjusting a single parameter while maintaining the others constant. This approach circumvents the need for multivariable optimization by implementing design space exploration. Further, to elucidate the relationships between different design elements and their impact on PTE based on Table I, a visual representation is provided in  Fig.~\ref{variable_sweep} indicating the optimal configurations for enhanced efficiency.

Fig.~\ref{variable_sweep}(a) illustrates that increasing $d_{h}$ from 0.1 mm to 14 mm reduces peak PTE and shifts peak frequencies downward. Notably, depths under 2 mm achieve PTEs over 27\% within the 1.4-1.5 GHz range, whereas deeper placements exhibit lower efficiencies, especially evident at 9 mm with minimal PTE around 6.5\%. Shallower implants excel at higher frequencies, contrary to deeper ones, which perform better below 1 GHz. In Fig.~\ref{variable_sweep}(b), variations in $d_{v}$ up to 10 mm result in a decline in peak PTE at 1.5 GHz without influencing the resonant frequency, leading to less distinct resonance characteristics as the distance increases.
Fig.~\ref{variable_sweep}\textcolor{black}{(d)} examines the impact of $t_{TX}$, revealing a relatively minor influence on PTE, with values over 22\% for thicknesses under 9 mm. Further insights from Fig.~\ref{variable_sweep}\textcolor{black}{(c)} and Fig.~\ref{variable_sweep}(e) indicate that increasing $d_{TX}$ from 20 mm to 50 mm and $r_{RX}$ from 0.5 mm to 3 mm shifts peak PTEs toward a stable resonant frequency of 1.5 GHz but reduces PTE magnitudes. This denotes a $>$ 10\% PTE for variable $d_{TX}$ and $>$ 19\% PTE for varying $r_{RX}$ respectively, within the mentioned range at 1.5 GHz. Lastly, Fig.~\ref{variable_sweep}(f) reveals how varying $h_{implant}$ affects PTE while maintaining a constant cross-sectional area. Lengths over 10 mm keep the resonant frequency at 1.5 GHz with PTEs above 14.5\%, whereas shorter lengths shift the frequency to 2.5 GHz.
% with PTE ranging from 0.83\% to 14.53\%

Hence, the resonant frequency remains stable unless the implant depth is varied. With other parameter adjustments, only the magnitude of PTE is affected. This underscores the critical balance among electrode dimensions, implant depth, and operational frequencies, emphasizing a trade-off in design. 

\subsection{Implant Misalignment Performance}
\begin{figure}[t!]
\centerline{\includegraphics[width=0.85\columnwidth]{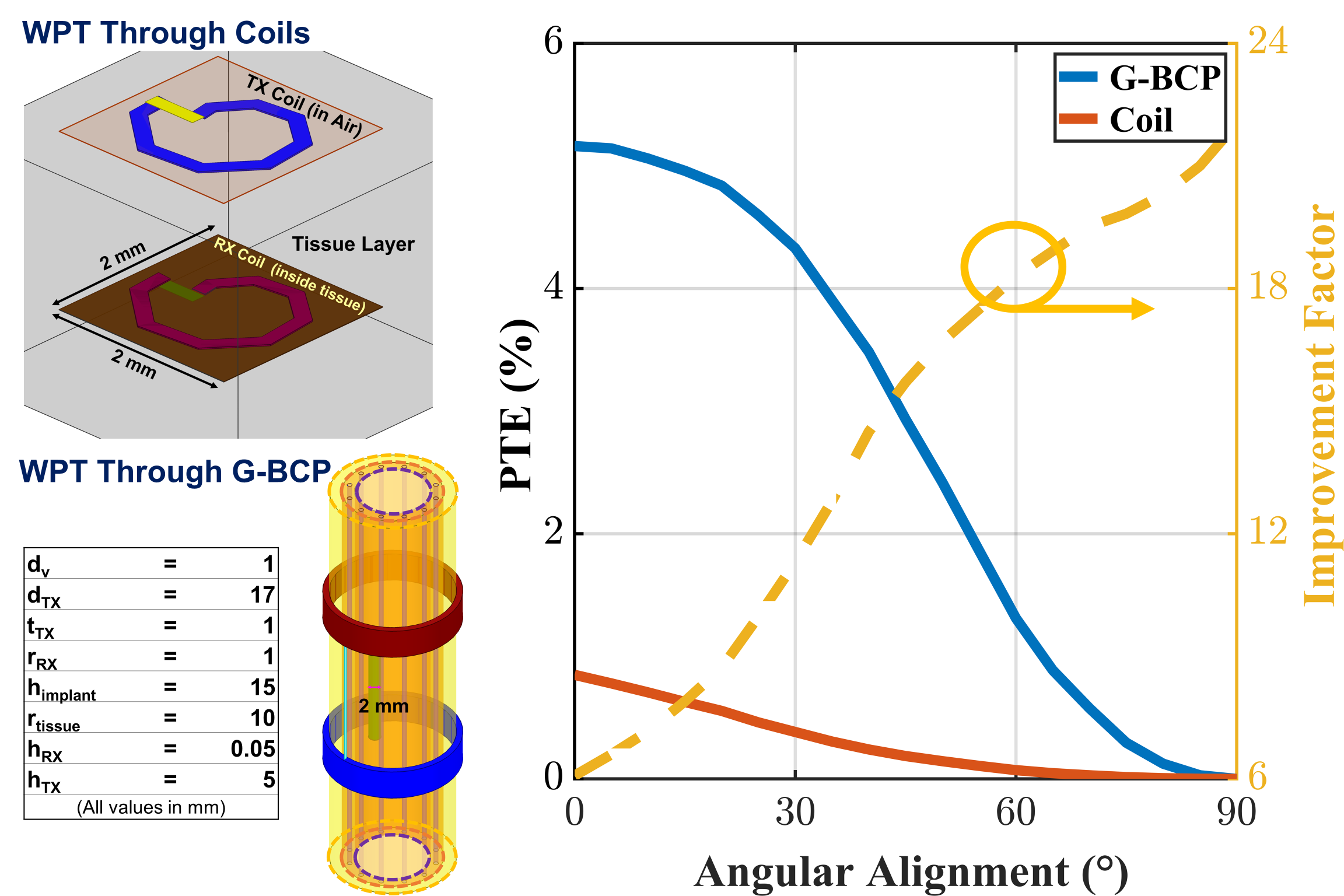}}
\vspace{-3mm}
\caption{Effect of angular misalignment of the implant at 0.6 GHz: performance of G-BCP in the worst case scenario vs. performance of resonant inductive coils with the comparable area. At an implant depth of 9 mm, the G-BCP surpasses coils with $>$ 6$\times$ improvement during perfect alignment and even $\approx$21$\times$ improvement for $85^{\circ}$ misalignment}
\label{bcpalign}
\vspace{-3mm}
\end{figure}

Practical considerations often result in misalignment during implant placement. Consequently, assessing how PTE responds to different alignment scenarios becomes crucial for the model developed thus far. In this context, Fig.~\ref{bcpalign} presents a case where the implant is embedded at a depth of 9 mm and undergoes rotation from $0^{\circ}$ to $90^{\circ}$ denoted by angular alignment, covering all possible orientations. Notably, it is observed that the resonant frequency remains constant at 0.6 GHz across all orientations, although PTE diminishes as misalignment increases. Remarkably, even when PTE falls below 1\% at rotation angles exceeding $70^{\circ}$, this performance is still superior compared to other WPT methodologies. As an example, Fig.~\ref{bcpalign} also illustrates the performance of a coil under various misalignments at 0.6 GHz. The cross-sectional area of the RX coils is kept at 2mm $\times$ 2mm compared to $\pi$ mm$^{2}$ area of the proposed G-BCP implant. Performance comparison with coils is conducted as I-WPT is recognized as the most effective and prevalent method among various near-field WPT techniques, according to section I. Moreover, the improvement in PTE elevates from 6$\times$ to 22$\times$ as the angular alignment increases from $0^{\circ}$ to $90^{\circ}$, which validates the robust resilience of the G-BCP approach to implant misalignments.

\begin{figure}[t]
\centerline{\includegraphics[width=0.9\columnwidth]{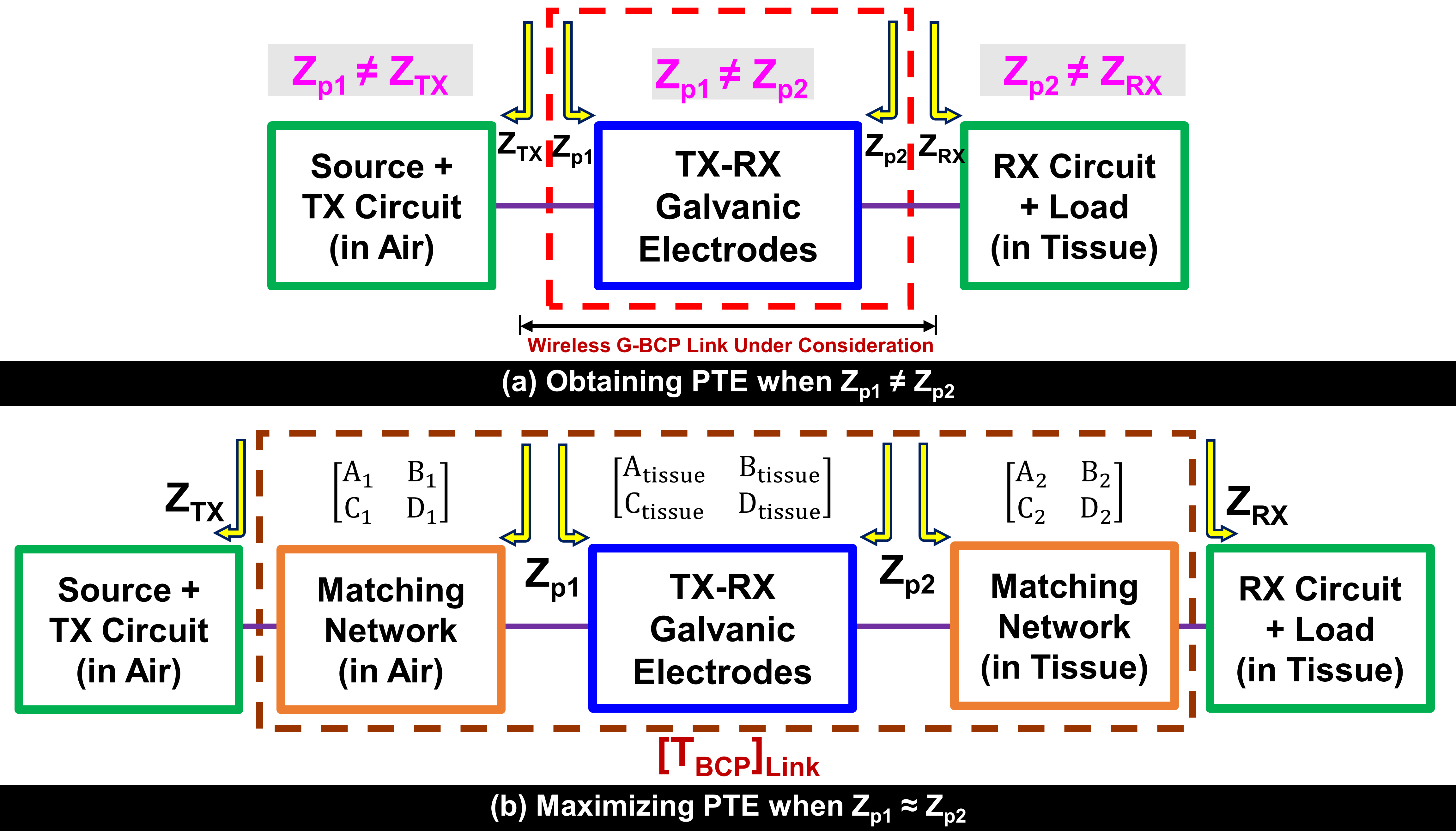}}
\vspace{-3mm}
\caption{PTE estimation based on port impedances looking into the source side and load side: (a) for $Z_{p1}\neq Z_{p2}$ (b) for $Z_{p1} = Z_{p2}$. To maximize PTE from the TX electrodes to RX electrodes through the tissue region, it is imperative to ensure the G-BCP link evaluates the same impedance, looking at both sides. Thus, it necessitates deploying IMNs at the TX and RX sides to minimize reflection at these reference planes and maximize power coupling.}
\label{maxpte}
\vspace{-3mm}
\end{figure}

\subsection{Maximizing PTE in the Wireless Link}
In the prior subsections, uniform port impedances ($Z_{p1}$, $Z_{p2}$) are assumed; however, this assumption does not necessarily ensure optimal PTE owing to discrepancies in the impedance as viewed from the source and load sides, as depicted in Fig.~\ref{maxpte}(a). This issue can be effectively addressed by incorporating an appropriate impedance matching network (IMN) on both sides, as illustrated in Fig.~\ref{maxpte}(b). The requisite analysis for this adjustment can be efficiently conducted using the ABCD parameters of each network component, as outlined in (1) and (2) \cite{pozar2011microwave}.
\vspace{-2mm}
\begin{equation}
[T_{BCP}]_{link} = 
\begin{bmatrix}
T_{1} & T_{2} \\ 
T_{3} & T_{4}
\end{bmatrix}
\end{equation}
\vspace{-2mm}
\begin{equation}
\begin{bmatrix}
T_{1} & T_{2} \\ 
T_{3} & T_{4} 
\end{bmatrix}=\begin{bmatrix}
A_{1} & B_{1} \\ 
C_{1} & D_{1} 
\end{bmatrix}
\begin{bmatrix}
A_{tissue} & B_{tissue} \\ 
C_{tissue} & D_{tissue} 
\end{bmatrix}
\begin{bmatrix}
A_{2} & B_{2} \\ 
C_{2} & D_{2} 
\end{bmatrix}
\end{equation}

Here, $[T_{BCP}]_{link}$ signifies the transmission performance in the G-BCP link. Also, the comprising ABCD matrices represent the IMN at the TX side, the transmission parameters of the BCP region between the TX and RX electrodes, and the IMN at the RX side, respectively, as also delineated in Fig.~\ref{maxpte}. This paves the way to analyze how maximum power can be coupled from TX electrodes to RX electrodes through the specified network components. Furthermore, converting $[T_{BCP}]_{link}$ to the corresponding S-matrix \cite{10182236,omi2021novel} results in deducing $S_{11_{BCP}}$ and $S_{22_{BCP}}$ shown in  (3) and (4), which represents the reflection parameters in the TX and RX side respectively.
\vspace{-3mm}
\begin{equation}
S_{11_{BCP}}=\frac{T_{1}Z_{p_{2}}+T_{2}-T_{3}Z_{p_{1}}Z_{p_{2}}-T_{4}Z_{p_{1}}}{T_{1}Z_{p_{2}}+T_{2}+T_{3}Z_{p_{1}}Z_{p_{2}}+T_{4}Z_{p_{1}}}\label{eq}
\end{equation}
%\vspace{-3mm}
\begin{equation}
S_{22_{BCP}}=\frac{-T_{1}Z_{p_{2}}+T_{2}-T_{3}Z_{p_{1}}Z_{p_{2}}+T_{4}Z_{p_{1}}}{T_{1}Z_{p_{2}}+T_{2}+T_{3}Z_{p_{1}}Z_{p_{2}}+T_{4}Z_{p_{1}}}\label{eq2}
\end{equation}
% \vspace{-4mm}

By setting $S_{11_{BCP}}$ = $S_{22_{BCP}}$ = 0, we can co-determine the IMN parameters at both sides. Thus, the systematic design process of G-BCP link is deemed complete, ensuring maximal PTE via the tissue medium.
\vspace{-1mm}
\subsection{Capability of Minimal Surgical Intervention}
The combination of small RX sizes and adjustable electrode-to-skin distances enhances the potential for injectable neural interfaces. Devices with a radius of 1-2 mm can be inserted through a 12-16G needle (inner diameter: 1.19-2.16 mm), facilitating deployment across various anatomical locations. This system pairs with minimally invasive electrodes that anchor to internal structures like muscles and connect via flexible leads, enabling precisely targeted medical interventions without open surgery \cite{hernandez2012minimally}. Such a design minimizes immune and foreign body responses, promoting seamless integration into bodily tissues \cite{khalifa2021injectable}. Although some WPT strategies can support even smaller receivers, they have limitations in PTE and tissue safety. In contrast, our G-BCP system can adeptly balance efficient power transfer with superior patient comfort and facilitate non-damaging neurostimulation of deep nerve targets within the body that are presently inaccessible, opening new avenues for advanced neurological therapeutics.

\section{Results and Discussion}
\begin{figure}[t!]
\centerline{\includegraphics[width=1\columnwidth]{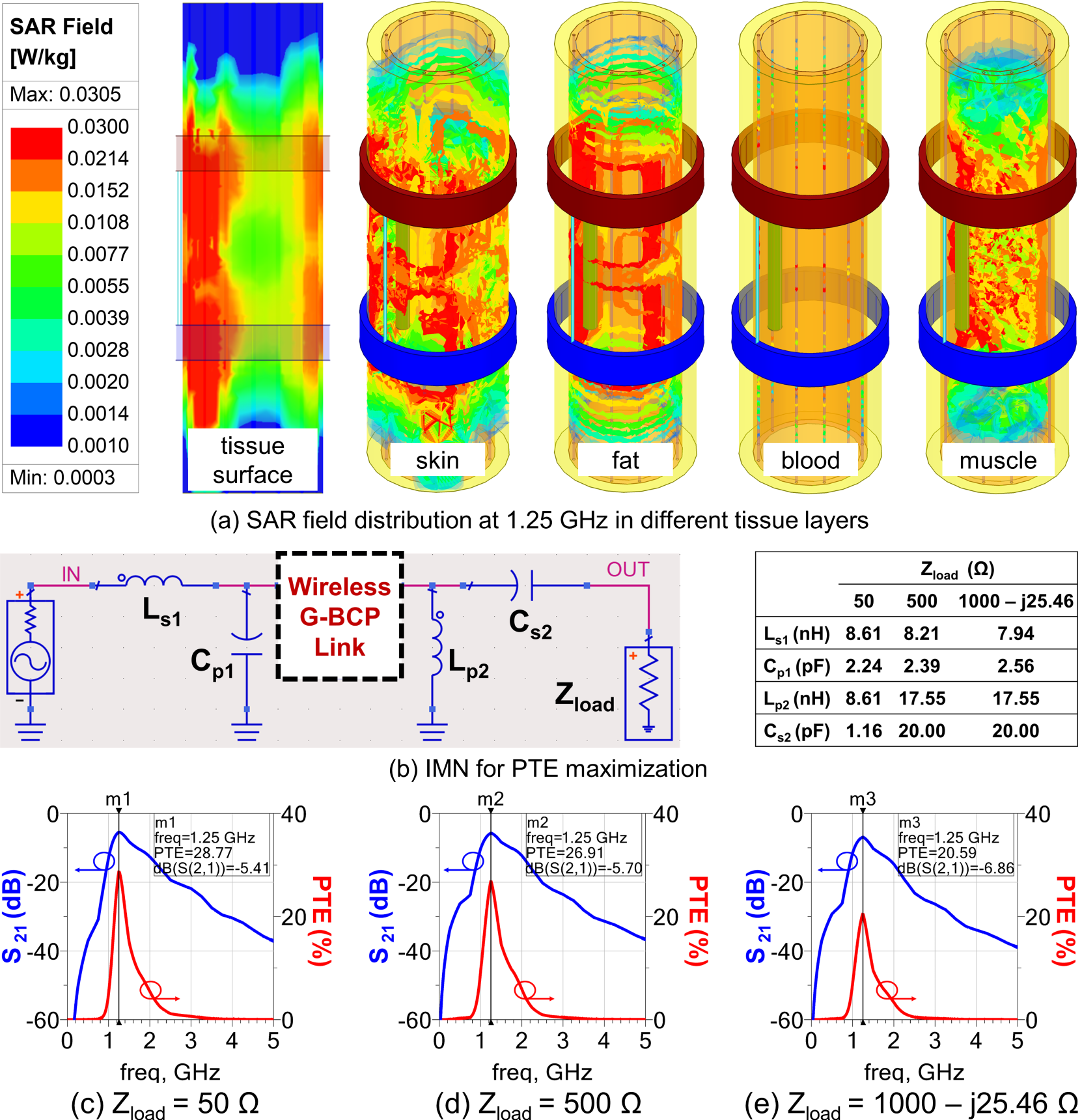}}
\vspace{-2mm}
\caption{Performance evaluation of the designed prototype at 1.25 GHz (a) E-field distribution and SAR estimate (b) IMN design for variable tissue load (c) PTE when $Z_{load} = 50 \Omega$ (d) PTE when $Z_{load} = 500 \Omega$ (e) PTE when $Z_{load} = 1000 -j25.46 \Omega$. After ensuring tissue safety from the SAR plot at 1.25 GHz, the graphs demonstrate the effectiveness of the L-section IMNs in optimizing PTE across varying load impedances. As impedance increases, PTE decreases, highlighting the importance of careful impedance matching in designing efficient wireless G-BCP systems.}
\label{prot}
\vspace{-5mm}
\end{figure}

To validate the proposed G-BCP system, a configuration was designed to maximize PTE at a 2 mm subcutaneous depth targeting the sciatic nerve of a rat. The optimized design parameters set the operational frequency at 1.25 GHz with specific dimensions: $d_{h}$ = 2mm, $d_{v}$ = 1mm, $t_{TX}$ = 1mm, $d_{TX}$ = 22mm, $r_{RX}$ = 1mm, and $h_{implant}$ = 20mm. To determine the tissue exposure to EM fields at 1.25 GHz, SAR field distribution was analyzed across the heterogeneous tissue layers as shown in Fig.~\ref{prot}(a). The peak simulated average SAR observed along the rat's leg was 0.031 W/kg, which is below the minimum thresholds prescribed by the International Commission on Non-Ionizing Radiation Protection (ICNIRP) for safety \cite{ICNIRP1998}. Specifically, the ICNIRP guidelines stipulate a lower limit of 0.08 W/kg for whole-body exposure and 4 W/kg for limb exposure. Thus, the thermal effects, even in the region with the higher SAR values, were considered minimal enough to ensure safety compliance, as corroborated by the SAR values. This estimate was cardinal for finalizing the device design as well.

Further, to maximize PTE at 1.25 GHZ, L-section IMNs are chosen due to their simpler implementation as depicted in Fig.~\ref{prot}(b). The specific values for L-section components ($L_{s1}$, $C_{p1}$, $L_{p2}$, $C_{s2}$) also vary with each load impedance, underscoring the need to adjust the matching network components to optimize performance for each specific load scenario. The IMN parameters are calculated based on (3) and (4). The final performance of the implemented G-BCP model under such variable loading is displayed in Fig.~\ref{prot}(c, d, e) as well. The PTE shifts from 28.8\% to 20.6\% when the \textcolor{black}{receiver} load impedance is varied from 50$\Omega$ to 1000 - j25.46$\Omega$ (assuming a resistance of 1000$\Omega$ plus a capacitance of 5pF at 1.25 GHz). Overall, the load impedance significantly affects the PTE, where lower impedance values are more efficiently matched with the circuit, resulting in higher PTE, which is a direct consequence of how well the L-section matching networks can transform the source impedance to match the load impedance at the resonance frequency.

\section{Conclusion}
This paper highlights an efficient way of powering miniaturized wireless neuroimplants through G-BCP. This system not only facilitates minimally invasive procedures as well as patient-centric at-home monitoring and treatments, but also demonstrates increased resilience against implant misalignments during bodily movements, as observed from the simulation. Further, the detailed analysis of the EM simulated results of the proposed G-BCP model combined with the IMN circuits suggests a more reliable and directed power delivery mechanism, setting new standards for WPT in the biomedical domain and significantly expanding the possibilities for complex neuroprosthetic applications. As future work, the EM prototypes will be fabricated and measured further to validate the optimal PTE at diverse load conditions plus WNMD alignments, and simultaneously, the implementation of pertinent TRX circuits for power delivery to the specific node will be investigated.

\bibliographystyle{IEEEtran}

% \bibliography{IEEEabrv,IEEEexample}

\bibliography{conference_101719}

\end{document}